\documentclass[runningheads]{svmult}

\usepackage{makeidx}   % allows index generation
\usepackage{graphicx}  % standard LaTeX graphics tool
\usepackage{subeqnar}  % subnumbers individual equations
\usepackage{multicol}  % used for the two-column index
\usepackage{physprbb}  % modified textarea for proceedings,
\makeindex             % used for the subject index
\begin{document}
\title*{Deep Near-Infrared Imaging of Galaxies at z $>$ 2 with Subaru}
\toctitle{Color and Magnitude Distribution in HDF-N and the field of 53W002}
% allows explicit linebreak for the table of content
%
%
\titlerunning{Deep NIR Imaging with Subaru}
% allows abbreviation of title, if the full title is too long
% to fit in the running head
%
\author{Toru Yamada\inst{1}
\and Masaru Kajisawa\inst{1}
}
\authorrunning{Yamada, T. and Kajisawa, M.}
% if there are more than two authors,
% please abbreviate author list for running head
%
%
\institute{National Astronomical Observatory of Japan, 
2-21-1, Osawa, Mitaka, Tokyo 181-8588
 }

\maketitle              % typesets the title of the contribution
\begin{abstract}
 We present the results of the deep NIR imaging observations using the Subaru telescope. At the Hubble Deep Field North, we obtained the new and deepest $K^\prime$-band image (10 hours) to study the rest-frame optical morphology of galaxies at z$> 3$ and the stellar mass distribution of galaxies at $2 < z < 4.5$. We also study the rest-frame optical properties of the 'building blocks' in the field of 53W002 and overall color distribution there. We then combined these data to compare the color and magnitude distribution in these two fields. The opt-NIR color-magnitude distributions in the two fields looks very similar and we report the conspicuous color change below $K_{AB} =22$ and discuss the cause of the feature.
\end{abstract}

\section{Extremely Deep NIR Imaging at the Hubble Deep Field North}

 While there exist quite deep and high-resolution WFPC2 and NICMOS images at the HDF-N, it has been desired to have the ground-base $K$-band data at the field with comparable depth and image quality in order to study more rest-frame optical properties of high-redshift galaxies. For the purpose, we obtained the very deep $K^\prime$-band image using the Subaru telescope equipped with a NIR camera, CISCO. With about 10 hours of net integration under the condition of 0.3-0.6 arcsec seeing, we reached to the depth of $K_{AB} \sim 25.5$ as the peak of the counts of the detected galaxies.

 Using the data, combined with the archived HST data taken in the WFPC2 $U_{300}B_{450}V_{606}I_{814}$ as well as NICMOS $J_{110}H_{160}$ bands, we studied the (i) rest-frame optical morphology of the galaxies at $z > 3$, (2) stellar mass distribution of high-redshift galaxies, and (3) correlation between galaxy properties with the stellar mass. 

 Figure 1 shows the Subaru $K^\prime$-band image together with the HST $I_{814}$-band images. There is no clear Hubble sequence for galaxies at $z \sim 3$. The morphological feature seen in the HST WFPC2 images are well recognized in the Subaru $K^\prime$-band image.

 Figure 2 shows the obtained stellar mass distribution for the $K^\prime$-band selected galaxies between $z=1.9$ and 4.5. This is obtained by fitting the HST $UBVIJH$ and Subaru $K^\prime$-band photometric data by the stellar evolutionary synthesis models (GISSEL96; Buruzual and Charlot 1993) changing the star-formation history, IMF, age, extinction, etc (Papovich et al. 2001, Shapley et al. 2001). The filled circles show the galaxies with spectroscopic redshift and those only with photometric redshift are by the open circles. We note that the  stellar mass of the HDF-N $K^\prime$-selected galaxies are typically very small, in the range between $10^9$-$10^{10}$ $M_\odot$.

 In Figure 3, we present the possible correlation between stellar-mass
 and the rest-frame $U-V$ color for the galaxies. There are few massive
 (in stellar mass) galaxies with t
he bluest color and the massive galaxies tend to have redder $U-V$ color. If the rest $U-V$ color represents the age difference, then massive objects tend to have rather old average age, which may imply that the star-formation in these galaxies have been occurred rather successively, may be through the merging or assembly process.

\begin{figure}
%\begin{center}
%\includegraphics[width=1.\textwidth]{yamadaF1.eps}
%\end{center}
\caption[]{The rest-frame optical morphology of galaxies at z=3. We compare the WFPC2 raw (left) and seeing-convolved (middle) $I_{814}$ images with CISCO $K^\prime$ ones.}
\label{eps1}
\end{figure}

\begin{figure}
%\begin{center}
%\includegraphics[width=1.\textwidth]{yamadaF2.eps}
%\end{center}
\caption{Stellar mass distribution of the $K^\prime$-selected galaxies
 at z=1.9-4.5 in the HDF-N}
\label{partfig3}
\end{figure}

\begin{figure}
%\begin{center}
%\includegraphics[width=1.\textwidth]{yamadaF3.eps}
%\end{center}
\caption{Correlation between the obtained stellar mass and rest-frame $U-V$ color of the galaxies at z$\sim$2.}
\label{labelfig3}
\end{figure}

\section{The field of the Radio Galaxy 53W002 at z=2.4}

 We also obtained moderately deep $J$ and $K^\prime$-band images at the field of a radio galaxy 53W002 at z=2.4 where Pascarelle et al. (1996) discovered a dozen of emission-line objects and candidates using the HST WFPC2 intermediate-band image. Keel et al. (1999) also discovered that the field lies in the high-density region of strong emission line objects from the ground-base narrow-band images. We studied (i) the rest-frame optical properties of the emission-line selected objects and (ii) surface density of old quiescent galaxies at the redshift in the field using the Subaru data combined with the HST WFPC2 and NICMOS archive data. The detailed results are presented in Yamada et al. (2001). In brief summary, many of the emission-line objects are faint in NIR wavelength and so they are intrinsically compact and small objects dominated by the on-going star formation, and there are few developed quiescent galaxies that are older than $\sim 1$ Gyr in the field. The detailed discussion about the mass of the radio galaxy 53W002 itself based on the Subaru OHS and CISCO spectra is presented in Motohara et al. (2001).

 Figure 4 shows the $I-K^\prime$ color-magnitude diagram of the $K^\prime$-selected galaxies in the field of 53W002 (filled circles). We also noticed a somewhat conspicuous change of the color distribution at $K^\prime_{AB} \sim 22$. While the brighter objects show the colors distributed over the expected color rang for old and young galaxies at $z < 1$, below $K^\prime_{AB} \sim 22$, most of the galaxies have blue colors ($I_{814}-K^\prime_{AB} < 1.5$ ) and a small fraction have further red colors ($I_{814}-K^\prime_{AB} > 2.5$ ). There are few galaxies with intermediate color range and thus there is a conspicuous color 'gap' or 'void' in the diagram. This trend is also seen in the $J-K^\prime$ vs $K$ diagram (Yamada et al. 2001).

\section{Comparison of the Color-Magnitude Diagram in the Two Field}

 Then we compare the color-magnitude distribution on these two fields. The data for the objects in HDF-N are shown by the open squares in Figure 4, and strikingly, the color-magnitude distributions in the two fields are quite similar, although faint red population is not seen in HDF-N. It suggests that this kind of color-magnitude distribution is a typical aspect of the universe if not an average.

 What causes the conspicuous 'bluing' of the $I-K$ colors? Similar trend has in fact been reported by previous authors (Cowie et al. 1995, McCraken et al. 2000, Gardner 1995, Sarraco et al. 2001, but see Sarraco et al. 1999). We found that at least two phenomena are responsible for the distribution, namely, (i) deficit of intrinsically faint ($< L*$) red galaxies at intermediate redshift (z > 0.5-1), and (ii) deficit of red galaxies including luminous ones at $z > 1$. This can be well recognized by comparing the data with the passive evolutionary models for the color-magnitude relation in the Coma cluster (Kodama et al. 1998) that are also shown if figure 4. Three lines represent the tracks of passively-evolving galaxies that are formed at $z=4.6$ and have $M_V = -22, -20, -18$ at present epoch. The cosmology is $\Omega_0$=1, $\Omega_m$=0.3, $\Omega_{\lambda}$=0.7, and $H_0$=70 km s$^{-1}$ Mpc$^{-1}$.  
 
 For the galaxies in HDF-N, we have the data of spectroscopic and photometric redshift. We can see clear deficit of faint red galaxies for those between $z=0.5$ and $z=1$ as shown in Figure 5. This is quite puzzling and interesting phenomena, since it cannot be explained by simple field-to-field variation effects because there are a certain number of luminous red galaxies at the same redshift. The number density of low-mass objects are expected to be larger than that of the luminous objects since the local luminosity function have exponential cut-off at the luminous end at any band. Our result may imply that we rarely see the less massive galaxies that are dominated by old-stellar population at the intermediate redshift where old massive galaxies exist. It may be related to the negative faint-end of the local E/S0 luminosity function.

 Kajisawa and Yamada (2001) investigated a complete volume-limited sample of galaxies with $M_V < -20$ up to z=2 in the HDF-N. Based on their careful morphological classification as well as evaluation of the possible uncertainty in photometric redshift, they concluded that the number density of early-type galaxies in HDF-N have very conspicuous decrease above z=1. So the deficit of red galaxies at $z>1$ seen in Figure 4 may not be surprising although we still do not fully understand why such rapid change of the number density is observed at $z \sim 1$ after considering the possible effects of field-to-field variation and dust extinction, etc.  We are planning to conduct more extended deep NIR and optical survey with Subaru to fully resolve these questions.

\vspace{0.3cm}

 We thank Ichi Tanaka and Kentaro Motohara for useful discussions and help in observations. We also thank Taddy Kodama for useful suggestions and providing us the galaxy evolutionary synthesis models.

\begin{figure}
\begin{center}
\end{center}
\caption{Color-maginitude diagram for the $K^\prime$-selected objects in the 53W002 field (filled dots) and HDF-N (open squares). Passive evolution track of the Coma metallicity sequence model in Kodama et al. (1998) is also plotted. The formation epoch is z=4.6 and their present absolute magnitude values are $M_V = -22$, $-20$, and $-18$, respectively.}
\label{labelfig4}
\end{figure}

\begin{figure}
\begin{center}
\end{center}
\caption{Color-magnitude diagram for the $K^\prime$-selected galaxies at
 z=0.5 - 1.0 in 
 the HDF-N.}
\label{labelfig5}
\end{figure}

%INDEX%%%%%%%%%%%%%%%%%%%%%%%%%%%%%%%%%%%%%%%%%%%%%%%%%%%%%%%%%%%%%%%
% Please check with the editor of your book whether he plans to
% include a "mutual" subject index - if so, please code your entries
% in the standard syntax. For your own purposes you may print your
% "personal" index by using the following commands:
%
%\clearpage
%\addcontentsline{toc}{section}{Index}
%\flushbottom
%\printindex
%%%%%%%%%%%%%%%%%%%%%%%%%%%%%%%%%%%%%%%%%%%%%%%%%%%%%%%%%%%%%%%%%%%%%

\end{document}